\documentclass[preprint,showpacs,preprintnumbers,amsmath,amssymb,superscriptaddress]{revtex4}
\usepackage{graphicx,bm}

\newcommand{\Sec}[1]{Sec.\,\ref{#1}}

\newcommand{\nl}{\nonumber \\}

\newcommand{\be}{\begin{equation}}
\newcommand{\ee}{\end{equation}}
\newcommand{\bea}{\begin{eqnarray}}
\newcommand{\eea}{\end{eqnarray}}
\newcommand{\bsube}{\begin{subequations}}
\newcommand{\esube}{\end{subequations}}
\newcommand{\Fig}[1]{Fig.\,\ref{#1}}
\newcommand{\Eq}[1]{Eq.\,(\ref{#1})}

\newcommand{\alf}{\alpha}
\newcommand{\sgm}{\sigma}
\newcommand{\omg}{\omega}
\newcommand{\Omg}{\Omega}
\newcommand{\Gam}{\Gamma}
\newcommand{\gam}{\gamma}
\newcommand{\vpl}{\varepsilon}

\newcommand{\upa}{\uparrow}
\newcommand{\dwa}{\downarrow}
\newcommand{\la}{\langle}
\newcommand{\ra}{\rangle}
\newcommand{\dg}{\dagger}

\newcommand{\mb}{\mbox}

\newcommand{\tL}{\mbox{\tiny L}}
\newcommand{\tR}{\mbox{\tiny R}}
\newcommand{\rhoB}{\rho_{_{\rm B}}}
\newcommand{\GamL}{\Gamma_{\rm L}}
\newcommand{\GamR}{\Gamma_{\rm R}}

\newcommand{\rms}{{\rm s}}
\newcommand{\rmd}{{\rm d}}
\newcommand{\Lup}{{_{{\rm L}\!\uparrow}}}
\newcommand{\Ldw}{{_{{\rm L}\!\downarrow}}}
\newcommand{\Rup}{{_{{\rm R}\!\uparrow}}}
\newcommand{\Rdw}{{_{{\rm R}\!\downarrow}}}

\newcommand{\mtr}{
 {(\begin{tiny}\begin{array}{c}
 \!\!n_{{\rm L}\!\upa},
 n_{{\rm R}\!\upa}\!\!\\
 \!\!n_{{\rm L}\!\dwa},
 n_{{\rm R}\!\dwa}\!\!
 \end{array}\end{tiny})}}
 \newcommand{\mtrLA}{
 {(\!\begin{tiny}\begin{array}{c}
 \!\!n_{{\rm L}\!\upa}\!+\!1,
 n_{{\rm R}\!\upa}\!\!\\
 \!\!n_{{\rm L}\!\dwa},
 n_{{\rm R}\!\dwa}\!\!
 \end{array}\end{tiny}\!)}}
 \newcommand{\mtrLB}{
 {(\!\begin{tiny}\begin{array}{c}
 \!\!n_{{\rm L}\!\upa},
 n_{{\rm R}\!\upa}\!\!\\
 \!\!n_{{\rm L}\!\dwa}\!+\!1,
 n_{{\rm R}\!\dwa}\!\!
 \end{array}\end{tiny}\!)}}
 \newcommand{\mtrLC}{
 {(\!\begin{tiny}\begin{array}{c}
 \!\!n_{{\rm L}\!\upa}\!-\!1,
 n_{{\rm R}\!\upa}\!\!\\
 \!\!n_{{\rm L}\!\dwa},
 n_{{\rm R}\!\dwa}\!\!
 \end{array}\end{tiny}\!)}}
 \newcommand{\mtrLD}{
 {(\!\begin{tiny}\begin{array}{c}
 \!\!n_{{\rm L}\!\upa},
 n_{{\rm R}\!\upa}\!\!\\
 \!\!n_{{\rm L}\!\dwa}\!-\!1,
 n_{{\rm R}\!\dwa}\!\!
 \end{array}\end{tiny}\!)}}
 \newcommand{\mtrRA}{
 {(\!\begin{tiny}\begin{array}{c}
 \!\!n_{{\rm L}\!\upa},
 n_{{\rm R}\!\upa}\!+\!1\!\!\\
 \!\!n_{{\rm L}\!\dwa},
 n_{{\rm R}\!\dwa}\!\!
 \end{array}\end{tiny}\!)}}
 \newcommand{\mtrRB}{
 {(\!\begin{tiny}\begin{array}{c}
 \!\!n_{{\rm L}\!\upa},
 n_{{\rm R}\!\upa}\!\!\\
 \!\!n_{{\rm L}\!\dwa},
 n_{{\rm R}\!\dwa}\!+\!1\!\!
 \end{array}\end{tiny}\!)}}
 \newcommand{\mtrRC}{
 {(\!\begin{tiny}\begin{array}{c}
 \!\!n_{{\rm L}\!\upa},
 n_{{\rm R}\!\upa}\!-\!1\!\!\\
 \!\!n_{{\rm L}\!\dwa},
 n_{{\rm R}\!\dwa}\!\!
 \end{array}\end{tiny}\!)}}
 \newcommand{\mtrRD}{
 {(\!\begin{tiny}\begin{array}{c}
 \!\!n_{{\rm L}\!\upa},
 n_{{\rm R}\!\upa}\!\!\\
 \!\!n_{{\rm L}\!\dwa},
 n_{{\rm R}\!\dwa}\!-\!1\!\!
 \end{array}\end{tiny}\!)}}

\begin{document}

 \title{Spin-dependent current noises in transport
 through coupled quantum dots}

 \author{JunYan Luo}
 \affiliation{State Key Laboratory for
 Superlattices and Microstructures,
 Institute of Semiconductors,
 Chinese Academy of Sciences,
 P.O.~Box 912, Beijing 100083, China}
 \affiliation{Department of Chemistry,
 Hong Kong University of Science and
 Technology, Kowloon, Hong Kong}
 \email{firstluo@semi.ac.cn}

 \author{Xin-Qi Li}
 \affiliation{State Key Laboratory for
 Superlattices and Microstructures,
 Institute of Semiconductors,
 Chinese Academy of Sciences,
 P.O.~Box 912, Beijing 100083, China}
 \affiliation{Department of Chemistry,
 Hong Kong University of Science and
 Technology, Kowloon, Hong Kong}

 \author{YiJing Yan}
 \affiliation{Department of Chemistry,
 Hong Kong University of Science and
 Technology, Kowloon, Hong Kong}

\begin{abstract}
Current fluctuations can provide additional insight into quantum
transport in mesoscopic systems. %%
The present work is carried out for the fluctuation properties of
transport through a pair of coupled quantum dots which are connected
with ferromagnetic electrodes. Based on an efficient particle-number
resolved master equation approach, we are concerned with not only
fluctuations of the total charge and spin currents, but also of each
individual spin-dependent component.
%%%%
As a result of competition among the spin polarization, Coulomb
interaction, and dot-dot tunnel coupling, rich behaviors are found
for the self- and mutual-correlation functions of the spin-dependent
currents.
\end{abstract}


\maketitle

\section{Introduction}

 Noise of transport current through mesoscopic
 system can provide useful information
 beyond the average current \cite{Bla001,Naz03}.
 Typical examples include
 quantum shuttles \cite{Ble04159,LaH0474,%
 Nov04248302,Fli05411},
% ,Fli04205334,Rod05085324}
 spin valves \cite{Bra0601366,Gur05205341,%
 Bel04140407},
 double-dot
 structures \cite{Ela02289,Agh06195323,%
 Sun9910748,Agu04206601},
 as well as qubit
 measurement \cite{Mak01357,Li05066803,%
 Gur03066801,Sta04136802}.
%%%%
 These studies were focused largely on the total
 charge current.
 A relatively less exploited subject, but
now gaining increasing attention,
 is the spin-dependent current
 fluctuations \cite{Wan04153301,Don05066601,%
 Dju06115327,Sau04106601}.
%%%%%
 A number of interesting examples are listed as follows:
 (i) It was shown in \cite{Wan04153301}
 that the shot noise of spin current is closely
 related to the spin unit of quasiparticles.
%%%%%
(ii) In the absence of charge current, the
 magnetically pumped spin current noise through
 quantum dots in the Coulomb
 blockade regime can serve as sensitive
 probe to spin decoherence \cite{Don05066601}.
%%%%%
(iii) For a quantum dot strongly coupled to
 cavity field, shot noise of the spin current
 exhibits clear signature of the discrete
 nature of the photon states \cite{Dju06115327}.
%%%%%
 (iv) Spin-dependent shot noise of unpolarized
 currents can be used to detect attractive
 or repulsive correlations \cite{Sau04106601}.

 In this work we study the spin-dependent
 current noises for transport through a pair
 of coupled quantum dots connected with
 ferromagnetic (FM) electrodes. Owing to the
 additional modulation of the dot-dot coupling,
 rich noise behaviors are found as a consequence
 of interplay between the Coulomb correlation
 and the spin polarization of the FM electrodes.
 Methodologically, we employ an efficient
 particle-number-resolved master equation approach
 \cite{Li05066803,Li05205304,Luo07085325}, with
 proper extension by including the spin degrees
 of freedom.
%%%
 This approach, which can be considered as a
 finite temperature and voltage extension of
 the ``$n$"-resolved quantum Bloch-type
 equation proposed by Gurvitz and Prager
 \cite{Gur9615932}, has the advantages
 in its simplicity of treating properly a broad
 range of dissipation, as well as its transparency
 in the involving dynamics and many-body
 interactions.

 The paper is organized as follows. We begin
 in the next section with the model
 setup of double dots connected with FM electrodes.
%%%
 We then present in \Sec{thsec3} a
 general formalism for both the spin-dependent
 currents and their noises.
%%%
 Numerical results are presented in sections
 \ref{thsec4} and \ref{thsec5}, and
 then followed by the conclusion in \Sec{thsec6}.

 \section{\label{thsec2}Model Description}

 \begin{figure}
 \includegraphics*[scale=1]{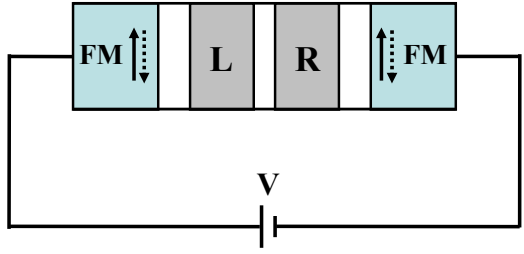}
 \caption{\label{GRA0}Schematic setup for
 transport through a pair of coupled quantum
 dots connected with ferromagnetic electrodes.}
 \end{figure}

 The system under study is schematically shown
 in \Fig{GRA0}.
 The left and right quantum dots are labeled by
 ``L'' and ``R'', respectively.
%%%%
 The total Hamiltonian contains three parts:
 \be
 \hat{H}=\hat{H}_{\rm B}+
 \hat{H}_{\rm S}+\hat{H}'.
 \ee
%%%%%
 $\hat{H}_{\rm B}=\sum_{\alf=
 {\rm L,R}}\sum_{k,\sgm}\vpl_{\alf k}
 c_{\alf k\sgm}^\dag c_{\alf k\sgm}$
 describe the left and right electrodes, which
 are magnetically polarized.
 $c_{\alf k\sgm}$ ($c_{\alf k\sgm}^\dag$)
 is the electron annihilation (creation) operator
 of the electrode $\alf=$ L or R,
 with spin $\sgm=\,\upa$ or $\dwa$.
%%%%%
 The FM electrodes are characterized by
 the spin-dependent density of states $g_{\alf\sgm}(\omg)$.
 Moreover, the densities of the states are assumed to
 be energy independent,
 i.e.\ $g_{\alf\sgm}(\omg)=g_{\alf\sgm}$.
 Then, the FM polarization of the electrodes is simply
 determined by
 a parameter $p_\alf=(g_{\alf\upa}
 -g_{\alf\dwa})/(g_{\alf\upa}+g_{\alf\dwa})$, which satisfies
 $-1\leqslant p_\alf\leqslant1$.

 The second Hamiltonian is for the coupled dots:
 \bea
 \hat{H}_{\rm S}
 &=&
   \sum_{\alf={\rm L,R}}\!\! E_\alf\hat{n}_\alf
  + U_0(\hat{n}_{\tL\upa}\hat{n}_{\tL\dwa}\!\!
  + \hat{n}_{\tR\upa}\hat{n}_{\tR\dwa})
  + U'\hat{n}_{\tL} \hat{n}_{\tR}
 \nl&&\quad
  + \Omg\sum_\sgm(d_{\tL\sgm}^\dag
    d_{\tR\sgm}\!+\!d_{\tR\sgm}^\dag
    d_{\tL\sgm}).
 \label{H-s}
 \eea
 Here $\hat{n}_{\alf\sgm}=d_{\alf\sgm}^\dag
 d_{\alf\sgm}$ and $\hat{n}_{\alf}=\sum_\sgm
 \hat{n}_{\alf\sgm}$,
 with $d_{\alf\sgm}$ ($d_{\alf\sgm}^\dag$) the electron
 annihilation (creation) operator in the quantum
 dot $\alf$ and with spin $\sgm$. Note that here $\alf$ labels
 the left and right quantum dots rather than the electrodes.
%%%%%
 Each quantum dot is assumed to have only one spin-degenerate
 energy level $E_{\rm L/R}$.
%%%%%
 $U_0$ and $U'$  are the intra-dot and inter-dots
 Coulomb interactions; and $\Omg$ is the coupling
 strength of the two dots.
%%%%%
 The typical value of intra-dot interaction $U_0$
 is estimated to be 4$\sim$5meV, and
 $U'\approx U_0/2$ \cite{Ono021313}.

 The third Hamiltonian describes the coupling of
 the central dots to the electrodes, in terms of
 $\hat{H}'=\sum_{\alf k\sgm}\big(t_{\alf k}
 c_{\alf k\sgm}^\dag d_{\alf\sgm}+{\rm H.c.}\big)$.
 Note that, due to spin polarization of the electrodes,
 the tunnel coupling strength would become spin-dependent,
 i.e.\ with $\Gam_{\alf\upa}=(1+p_\alf)\Gam_\alf/2$
 differing from
 $\Gam_{\alf\dwa}=(1-p_\alf)\Gam_\alf/2$,
 where
 $\Gam_\alf=2\pi\sum_k|t_{\alf k}|^2\delta
 (\vpl_{\alf k}-\omg)$ is the total
 coupling strength regardless the spin orientation,
 and is estimated
 to be $10^{9}\sim10^{10}$~s$^{-1}$ \cite{McC07056801}.

 In this work, corresponding to different bias
 voltages, we will \emph{effectively} consider
 the following three cases.
%%%%%
 Case (i): noninteracting (NINT) dots, i.e.\
 each dot can be doubly occupied.
%%%%%
 Case (ii): single-dot Coulomb blockade (SDCB).
 In this case double occupation on the same dot
 is prohibited, but each of the two dots can
 hold an electron.
%%%%%
 Case (iii): double-dot Coulomb blockade (DDCB).
 In this case, there is at most only one
 electron in the two dots.
%%%%%
 Experimentally, the above mentioned different
 cases can be achieved by appropriately adjusting
 the applied bias voltage with respective to the
 intradot and interdot charging
 energies \cite{Luo07085325}.

 \section{\label{thsec3} Formalism}

 In this section we outline the master equation
 approach for the calculation of the spin-dependent
 shot noise. The main point is to obtain the
 equation of motion for the \emph{conditional}
 reduced density matrix of the central quantum dots,
 which is conditioned not only on the transmitted
 electron numbers, but also on their spins.

 \subsection{Electron-number-resolved master equation}

 To achieve the description of spin-dependent
 transport, we will first extend
 the ``$n$''-resolved quantum master
 equation \cite{Li05066803,Li05205304,Luo07085325}
 to include spin degrees of freedom.
 Let us start with the reduced density
 matrix, defined as $\rho(t)\equiv
 \mb{Tr}_{\rm B}[\rho_{\rm T}(t)]$,
 i.e.\ tracing over the electrode states from the total density matrix.
 Assuming the tunneling Hamiltonian
 $\hat{H}'$ is weak, the second-order
 cumulant expansion leads to the master
 equation \cite{Yan982721},
 \be\label{cumm-expan}
 \dot{\rho}(t)=-i{\cal L}\rho(t)
 \!-\!\int_0^\infty\!\!\!d\tau\la
 {\cal L}'(t){\cal G}(t,\tau){\cal L}'
 (\tau){\cal G}^\dag(t,\tau)\ra\rho(t),
 \ee
 where the Liouvillian superoperators are
 defined as
 ${\cal L}A\equiv[\hat{H}_{\rm S}, A]$,
 ${\cal L}'A\equiv[\hat{H}',A]$, and ${\cal G}
 (t,\tau)A\equiv G(t,\tau)AG^\dag(t,\tau)$,
 with $G(t,\tau)$ the usual propagator
 associated with the system Hamiltonian
 $\hat{H}_{\rm S}$.
%%%%%

 To condition the master equation on the electron number
 transmitted \cite{Li05066803,Li05205304,Luo07085325},
 we decompose the entire Hilbert space
 ${\rm B}$ of the left and right electrodes into
 the sum of the subspaces,
 $\mb{B}^{(\!n_{\Lup}\!,n_{\Ldw}\!)}_{\rm L}
 \otimes
 \mb{B}^{(\!n_{\Rup}\!,n_{\Rdw}\!)}_{\rm R}$,
 where $n_{\alf\sgm}$ is the number of electrons with spin-$\sgm$
 arrived at the electrode $\alf$ ($\alf$=L, or R).
 Partially tracing over states in each subspace
 leads to the spin-resolved quantum master
 equation for the reduced density matrix,
 \bea\label{CQME}
 \dot{\rho}^{\mtr}\!\!\!
 &=&\!\!-i\mathcal{L}\rho^{\mtr}
 \!-\!\frac{1}{2}\bigg\{\!\sum_{\alf\sgm}
 \!\bigg[d^\dag_{\alf\sgm}
 A^{(\!-\!)}_{\alf\sgm}\rho^{\mtr}+
 \rho^{\mtr}A^{(\!+\!)}_{\alf\sgm}
 d^\dag_{\alf\sgm}
 \bigg]
 \nl
 &&\!-\bigg[
 d^{\dag}_{{\rm L}\upa}\rho^{\mtrLA}
 A^{(\!+\!)}_{{\rm L}\!\upa}\!
 +\!d^{\dag}_{{\rm L}\dwa}\rho^{\mtrLB}
 A^{(\!+\!)}_{{\rm L}\!\dwa}
 \!+\!A^{(\!-\!)}_{{\rm L}\!\upa}
 \rho^{\mtrLC}d^{\dag}_{{\rm L}\upa}
 \nl
 &&\!+A^{(\!-\!)}_{{\rm L}\dwa}\rho^{\mtrLD}
 d^{\dag}_{{\rm L}\dwa}
 +d^{\dag}_{{\rm R}\upa}\rho^{\mtrRA}\!
 A^{(\!+\!)}_{{\rm R}\!\upa}
 +d^{\dag}_{{\rm R}\dwa}\rho^{\mtrRB}\!
 A^{(\!+\!)}_{{\rm R}\!\dwa}
 \nl
 &&\!+A^{(\!-\!)}_{{\rm R}\!\upa}\rho^{\mtrRC}
 d^{\dag}_{{\rm R}\upa}
 \!+A^{(\!-\!)}_{{\rm R}\!\dwa}\rho^{\mtrRD}
 d^{\dag}_{{\rm R}\dwa}
 \,\bigg]+\mb{H.c.}\bigg\}.
 \eea
%%%%
 Here, $A_{\alf\sgm}^{(+)}
 \equiv\sum_{\sgm'}C_{\alf\sgm'\!\sgm}^{(+)}
 ({\cal L})d_{\alf\sgm'}$ and $A_{\alf\sgm}^{(-)}
 \equiv\sum_{\sgm'}C_{\alf\sgm\sgm'}^{(-)}(-{\cal L})$
 $d_{\alf\sgm'}$, with $C_{\alf\sgm\sgm'}^{(\pm)}
 (\pm {\cal L})=\int%^{\infty}_{-\infty}
 dt C_{\alf\sgm\sgm'}^{(\pm)}(t)
 e^{\pm i{\cal L}t}$. The involved bath
 correlation functions are defined as
 $C_{\alf\sgm\sgm'}^{(+)}(t-\tau)\equiv\la
 f_{\alf\sgm}^{\dg}(t)f_{\alf\sgm'}(\tau)\ra$
 and $C_{\alf\sgm\sgm'}^{(-)}(t-\tau)\equiv\la
 f_{\alf\sgm}(t)f^\dag_{\alf\sgm'}(\tau)\ra$,
 with $f_{\alf\sgm}=\sum_kt_{\alf \sgm}c_{\alf
 k\sgm}$ and $\la \cdots \ra\equiv
 \mb{Tr}_{\rm B} [(\cdots)\rhoB]$ standing
 for the usual meaning of thermal bath
 average.

 \subsection{Spin-dependent currents}

 With the knowledge of the above conditional
 state, the joint probability distribution
 function is obtained as
 $P[\mtr,t]\equiv\mb{Tr}\rho^\mtr$,
 where $\mb{Tr}(\cdots)$ denotes the trace
 over the system states.
 It allows us to evaluate the spin-$\sgm$ dependent
 current through the junction $\alf$ by
 \be
 I_\alf^\sgm=e\frac{d}{dt}
 \sum_{\!n_{\!\tL\!\upa\!},n_{\tL\!\dwa\!}}
 \sum_{\!n_{\!\tR\!\upa\!},n_{\tR\!\dwa\!}}
 \!\!n_{\alf\sgm}P\left[\mtr,t\right]
 =e\mb{Tr}\dot{N}_\alf^\sgm,
 \ee
 where $N_\alf^\sgm\equiv
 \sum_{\!n_{\!\tL\!\upa\!},n_{\tL\!\dwa\!},
 n_{\!\tR\!\upa\!},n_{\tR\!\dwa\!}}\!
 n_{\alf\sgm}\rho^{\mtr}$.
 Based on \Eq{CQME}, we obtain
 \bea\label{N-EOM}
 \frac{d}{dt}N_\alf^\sgm\!\!\!&=&\!\!
 -i\mathcal{L}N_\alf^\sgm
 -\frac{1}{2}\Big\{\!\sum_{\alf'\sgm'}\!
 \big[d^\dag_{\alf'\sgm'},
 A^{(-)}_{\alf'\sgm'}N_\alf^\sgm\!-\!
 N_\alf^\sgm A^{(+)}_{\alf'\sgm'}\big]
 \nl
 &&+\big[d^\dag_{\alf\sgm}\rho(t)
 A^{(+)}_{\alf\sgm}\!-\!A^{(-)}_{\alf\sgm}
 \rho(t)d^\dag_{\alf\sgm}\big]
 +\mb{H.c.}\Big\}.
 \eea
 This leads to
 \be\label{current}
 I_\alf^\sgm={\textstyle \frac{1}{2}}e\mb{Tr}
 \big\{\big[d_{\alf\sgm}^\dag
 A_{\alf\sgm}^{(-)}-A_{\alf\sgm}^{(+)}
 d_{\alf\sgm}^\dag\big]\rho(t)
 +\mb{H.c.}\big\},
 \ee
 where $\rho(t)$ is the unconditional
 density matrix that satisfies \Eq{cumm-expan}, or
 \be\label{QME}
 \dot{\rho}(t)=-i{\cal L}\rho(t)-\frac{1}{2}
 \sum_{\alf\sgm}\left\{\left[d_{\alf\sgm}^\dag,
 A^{(\!-\!)}_{\alf\sgm}\rho(t)
 \!-\!\rho(t)A^{(\!+\!)}_{\alf\sgm}\right]\!
 +\!\mb{H.c.}\right\}.
 \ee

 \subsection{Spin-dependent noises}

 Note that the total charge and spin
 currents through junction $\alf$ are
 $I_\alf^{\rm ch}=
 I_\alf^\upa\!+\!I_\alf^\dwa$ and
 $I_\alf^{\rm sp}=I_\alf^\upa\!-
 \!I_\alf^\dwa$. The total charge current
 noise and spin current noise can then be
 expressed as
 \bsube\label{Stot}
 \bea
  S_{\alf\alf'}^{\rm ch}
 &=&
   S_{\alf\alf'}^{\upa\upa}
  +S_{\alf\alf'}^{\dwa\dwa}
  +S_{\alf\alf'}^{\upa\dwa}
  +S_{\alf\alf'}^{\dwa\upa},
 \\
  S_{\alf\alf'}^{\rm sp}
 &=&
   S_{\alf\alf'}^{\upa\upa}
  +S_{\alf\alf'}^{\dwa\dwa}
  -S_{\alf\alf'}^{\upa\dwa}
  -S_{\alf\alf'}^{\dwa\upa},
 \eea
 \esube
 where the individual spin-dependent noise
 spectrum is defined as
 $S_{\alf\alf'}^{\sgm\sgm'}(\omg)\!\equiv\!
 \int_{-\infty}^\infty \!\!
 dt\,\cos(\omg t)
 \la\{\Delta I_\alf^\sgm(t),
 \Delta I_{\alf'}^{\sgm'}(0)\}\ra$,
 with $\Delta I_\alf^\sgm(t)\equiv
 I_\alf^\sgm(t)-\bar{I}_\alf^\sgm$.
 In the following, fluctuation between
 the same or opposite spin currents is
 referred to as self- or mutual-correlation
 shot noise, respectively.
%%%%%
 Based on the MacDonald's
 formula \cite{Mac62,Fli05411,Moz02161313},
 it follows that
 \be\label{Mac}
 S_{\alf\alf'}^{\sgm\sgm'}(\omg)=
 2\,\omg\!\int_{0}^{\infty}\!\!
 dt\sin(\omg t)
 \frac{d}{dt}\big[
 M_{\alf\alf'}^{\sgm\sgm'}(t)
 \!-\!\bar{I}_\alf^\sgm
 \bar{I}_{\alf'}^{\sgm'}t^2\big],
 \ee
 where $\bar{I}_\alf^\sgm$ is the stationary
 current obtained from \Eq{QME}, and
 $M_{\alf\alf'}^{\sgm\sgm'}(t)
 \equiv e^2\sum_{n_{\tL\!\upa\!},
 n_{\tL\!\dwa\!},n_{\tR\!\upa\!},
 n_{\tR\!\dwa\!}}\!n_{\alf\sgm}
 n_{\alf'\sgm'}P\big[\mtr,t\big]$.
 Using the spin-resolved master
 equation (\ref{CQME}), one finds
 \bea\label{M-t}
 \!\!\frac{d}{dt}
 M_{\alf\alf'}^{\sgm\sgm'}\!
 \!\!&=&\!\!
 {\textstyle \frac{1}{2}}e^2\mb{Tr}
 \big\{\big[d^\dag_{\alf\sgm}
 A^{(\!-\!)}_{\alf\sgm}\!
 -\!A^{(\!+\!)}_{\alf\sgm}
 d^\dag_{\alf\sgm}\big]
 N_{\alf'}^{\sgm'}\!(t)\!
 \nl
 &&\!\!\!+\big[
 d^\dag_{\alf'\!\sgm'}
 A^{(\!-\!)}_{\alf'\!\sgm'}
 \!-\!A^{(\!+\!)}_{\alf'\!\sgm'}
 d^\dag_{\alf'\!\sgm'}\!\big]
 N_{\alf}^{\sgm}(t)
 \nl
 &&\!\!\!+\big[
 d^\dag_{\alf\sgm}
 A^{(\!-\!)}_{\alf\sgm}
 \!+\!A^{(\!+\!)}_{\alf\sgm}
 d^\dag_{\alf\sgm}\!\big]
 \rho_{\rm st}\delta_{\alf\alf'}
 \delta_{\sgm\sgm'}
 \!\!+\!\mb{H.c.}\!\big\},
 \eea
 where $N_\alf^\sgm(t)$ can be found from
 \Eq{N-EOM}, and $\rho_{\rm st}$ is the
 stationary solution of the master equation
 (\ref{QME}).
%%%%%
 Specially, the zero-frequency spin-dependent
 noise which is of most interest
 reads \cite{Fli05411,Ela02289,Don05066601}
 \be
 S_{\alf\alf'}^{\sgm\sgm'}
 =2\frac{d}{dt}\big[M_{\alf\alf'}^{\sgm\sgm'}(t)
 -\bar{I}_\alf^\sgm\bar{I}_{\alf'}^{\sgm'}t^2
 \big]\big|_{t\rightarrow\infty}.
 \ee

 So far we have outlined a compact formalism
 for spin-dependent transport through
 mesoscopic systems.
%%%%%
 As will show in the following, this approach
 can be efficiently used to study the
 spin-dependent transport phenomena, say,
 not only the spin-dependent currents, but
 also their noise properties.
%%%%%%
 The major approximation involved in our
 formalism is the standard second-order
 Born approximation with respect to the
 perturbative expansion of the tunneling
 Hamiltonian.
%%%%
 Apparently, this approach is applicable
 under the Markov-type condition for relatively
 high temperatures, say, higher than the
 tunneling width \cite{Wun05205319}.
 In this case, the formalism can be safely applied
 for arbitrary bias voltages, including those
 with Fermi level of electrode in near
 resonance with the energy level of the
 central (transport) system.
%%%%
 Also, the present formalism is applicable to
 low temperature under large bias voltage, i.e.\
 the electrode's Fermi surface is further away
 from the central system's energy level than
 the level broadening.
%%%%%
 Similar statement holds also for the widely
 applied rate equation approach at zero
 temperature \cite{Gur9615932}.
%%%%
 Finally, at low temperature and under low bias
 voltage (near-resonance transport), an improved
 self-consistent Born approximation \cite{Cui06449}
 or exact hierarchical equations of motion
 approach \cite{Jin07134113} can be satisfactorily
 employed.

 \begin{table*}
 \caption{\label{table}Spin-dependent currents,
 as well as the total charge and
 spin currents for (i) NINT case, (ii) SDCB case,
 and (iii) DDCB case, respectively.
 Here, $\gam_0\equiv\GamL+\GamR$,
 $\gam_\rms\equiv2\GamL+\GamR$, and
 $\gam_\rmd\equiv4\GamL+\GamR$
 are the corresponding total effective
 tunneling rates for these three cases.}
 \begin{tabular}{cccc}\hline\hline
 Cases & (i) & (ii) & (iii) \\
 \hline
 $\bar{I}^{\upa}/e$ &
 $(1+p)\frac{\GamL\GamR\Omg^2/\gam_0}{\Omg^2
 +\frac{1}{4}(1+p)^2\GamL\GamR}$
&
 $\frac{1+p}{2}
 \frac{2\GamL\GamR\Omg^2/\gam_\rms}{\Omg^2+\frac{1}{2}
 (1+p^2\GamR/\gam_\rms)\GamL\GamR}$
&
 $\frac{1+p}{2}
 \frac{4\GamL\GamR\Omg^2/\gam_\rmd}{2\Omg^2+(1+p^2)
 (\GamL\GamR^2/\gam_\rmd)}$
\vspace{1ex}\\
 $\bar{I}^{\dwa}/e$ &
 $(1-p)\frac{\GamL\GamR\Omg^2/\gam_0}{\Omg^2
 +\frac{1}{4}(1-p)^2\GamL\GamR}$
&
 $\frac{1-p}{2}
 \frac{2\GamL\GamR\Omg^2/\gam_\rms}{\Omg^2+\frac{1}{2}
 (1+p^2\GamR/\gam_\rms)\GamL\GamR}$
&
 $\frac{1-p}{2}
 \frac{4\GamL\GamR\Omg^2/\gam_\rmd}{2\Omg^2+(1+p^2)
 (\GamL\GamR^2/\gam_\rmd)}$
\vspace{1ex}\\
 $\bar{I}^{\rm ch}/e$ &
 $\frac{8\GamL\GamR\Omg^2[(1-p^2)\GamL\GamR+4\Omg^2]/\gam_0}
 {[(1+p^2)\GamL\GamR+4\Omg^2]^2
   -4p^2\GamL^2\GamR^2}$
&
 $\frac{2\GamL\GamR\Omg^2/\gam_\rms}{\Omg^2+\frac{1}{2}
 (1+p^2\GamR/\gam_\rms)\GamL\GamR}$
&
 $\frac{4\GamL\GamR\Omg^2/\gam_\rmd}{2\Omg^2+(1+p^2)
 (\GamL\GamR^2/\gam_\rmd)}$
\vspace{1ex}\\
 $\bar{I}^{\rm sp}/e$ &
 $p\frac{8\GamL\GamR\Omg^2[(p^2-1)\GamL\GamR+4\Omg^2]/\gam_0}
 {[(1+p^2)\GamL\GamR+4\Omg^2]^2
 -4p^2\GamL^2\GamR^2}$
&
 $p\frac{2\GamL\GamR\Omg^2/\gam_\rms}{\Omg^2+\frac{1}{2}
 (1+p^2\GamR/\gam_\rms)\GamL\GamR}$
&
 $p\frac{4\GamL\GamR\Omg^2/\gam_\rmd}{2\Omg^2+(1+p^2)
 (\GamL\GamR^2/\gam_\rmd)}$\vspace{1ex}\\ \hline\hline
 \end{tabular}
 \end{table*}

 \section{\label{thsec4} Current characteristics}

 \begin{figure*}
 \includegraphics*[scale=0.8]{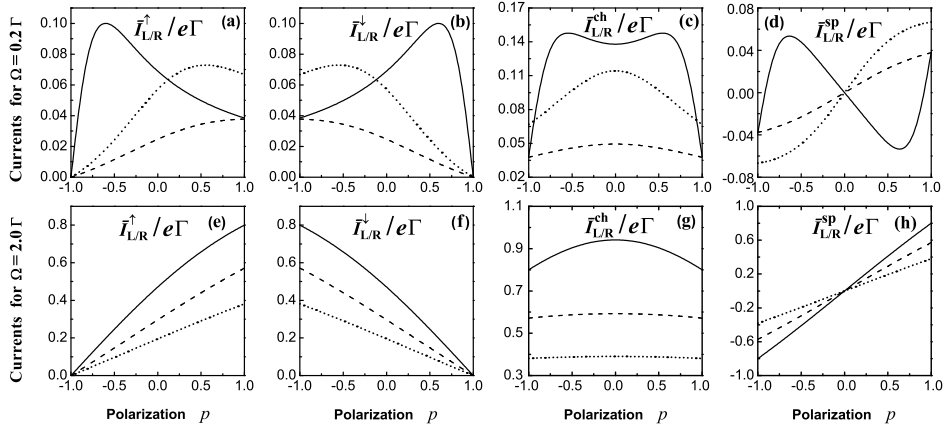}
 \caption{\label{GRA1}Currents versus spin
 polarization $p$ for (i) NINT
 case (solid curves), (ii) SDCB case (dashed
 curves), and (iii) DDCB case (dotted curves),
 respectively.
%%%%%
 Figures (a)-(d) show respectively, the spin-up,
 spin-down, total charge and total spin currents
 for $\Omg=0.2\Gam$, and those in (e)-(h) are
 the corresponding currents for $\Omg=2.0\Gam$.}
 \end{figure*}

 For simplicity, we assume the left and right
 electrodes have the same spin polarization
 $p_{\rm L}=p_{\rm R}=p$, and the two levels are in
 resonance, i.e.\ $E_{\rm L}=E_{\rm R}$. This can
 be achieved, in experiment, by applying
 appropriately gate voltages on the left and right
 dots (see, for instance, Ref.\ \cite{Pet052180}).
%%%%%
 Hereafter, we consider large bias voltage and
 low temperature in each case, which makes the
 Fermi functions relevant to the transport
 processes be either one or zero.
%%%%%
 Under this condition, electrons only transport
 in one direction, say, from left electrode to
 the right one.

%%%%
 Listed in table \ref{table} are the stationary
 spin-dependent currents, together with the total
 charge and spin currents.
%%%
 Numerical results are presented in
 \Fig{GRA1}, where we further assume symmetric
 coupling, i.e.\ $\GamL=\GamR=\Gam$.
%%%%%
 Here, the currents are plotted against the spin
 polarization under the condition of large bias
 voltage.
%%%%%
 At finite bias, the energy renormalization will
 be involved in the present formalism. It arises
 from the imaginary part of the rate \cite{Wun05205319},
 and can be included readily within our theory
 by following \cite{Yan05187,Xu029196}.
%%%%%
 The main features found here are summarized
 as follows:
%%%%%
 (i) the currents are suppressed in the
 presence of Coulomb interactions;
%%%%%
 (ii) there exists a symmetry relation between
 $\bar{I}^{\upa}$ and $\bar{I}^{\dwa}$, which
 makes us only need to consider one of them,
 e.g.\ $\bar{I}^{\upa}$;
%%%%%
 (iii) the spin-up currents ($\bar{I}^{\upa}$)
 and the total spin currents increase
 monotonically with $p$, while the charge
 currents reach their maxima at $p=0$.
%%%%%

 The results for the noninteracting and weakly
 coupled dots are shown by the solid curves in
 \Fig{GRA1}(a)-(d), which differ from the
 behaviors stated above in (iii).
%%%%%
 In this weak dot-dot coupling regime,
 the resultant large occupation probability of
 the left dot leads to simple expression for the
 currents.
%%%%%
 For instance, the spin-up current in this
 case can be well approximated by
 $\bar{I}^{\upa}\approx\Gam_{{\rm L}\upa}
 \rho_{{\rm L}\dwa}$ with $\rho_{{\rm L}\sgm}$
 the probability of the left dot being
 occupied by a spin-$\sgm$ electron.
 As the polarization of the FM electrodes is
 gradually altered from $p=-1$ to $p=1$,
 $\rho_{{\rm L}\dwa}$ decreases rapidly, while
 $\Gam_{{\rm L}\upa}$ increases linearly. Then,
 a turnover behavior for the spin-up current is
 formed as shown by the solid curve in
 \Fig{GRA1}(a).
%%%%%
 This is also the basic reason for the unique
 features of the charge and spin currents shown by
 the solid curves in \Fig{GRA1}(c) and (d).

 \begin{figure*}
 \includegraphics*[scale=0.9]{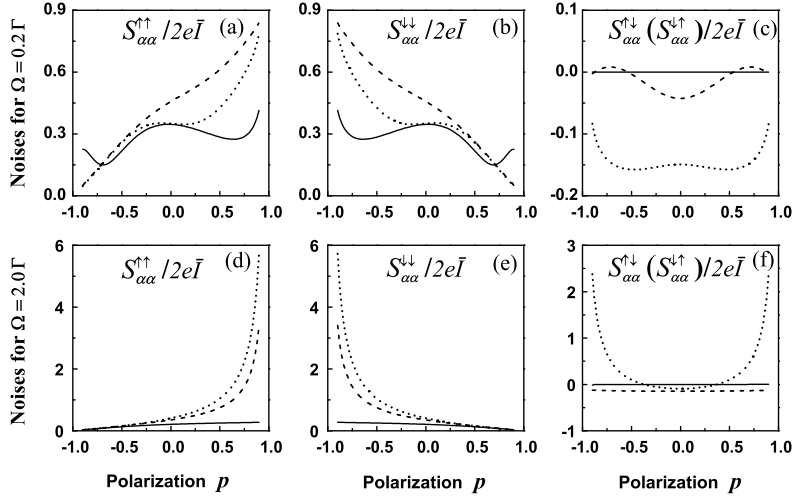}
 \caption{\label{GRA2}Individual spin-dependent
 noise for NINT case (solid curves),
 SDCB case (dashed curves), and DDCB case
 (dotted curves), respectively.
 Figures (a)-(c) are the noises for weak
 coupling ($\Omg=0.2\Gam$), and
 those for strong coupling ($\Omg=2.0\Gam$)
 are shown in (d)-(f).}
 \end{figure*}

 \section{\label{thsec5}Noise characteristics}

 In this section, instead of using the circuit
 noise \cite{Gur05205341,Agu04206601,Luo07085325},
 we will investigate various
 fluctuations of spin currents through left or
 right junction, i.e.\ the auto-correlations
 $S_{\alf\alf}^{\sgm\sgm'}$ with $\alf=$ L or R.
%%%%%
 For the cross-correlations, they simply satisfy
 the relation $S_{{\rm L}{\rm R}}^{\sgm\sgm'}
 =-S_{\alf\alf}^{\sgm\sgm'}$ in the present
 two-terminal case, as we have checked.
%%%%%
 It should be noted that for a three-terminal
 structure such a simple relation generally
 does not take place (see, for instances,
 Ref.\ \cite{Bag03085316,Cot04115315,Cot04206801}).

 \subsection{Self-correlation shot noise}

 We now turn to the noise characteristics,
 and first to the self-correlation shot noise.
%%%%%
 Note that there is a symmetry between
 $S^{\upa\upa}_{\alf\alf}$
 and $S^{\dwa\dwa}_{\alf\alf}$ with respect to
 the spin polarization, i.e., \Fig{GRA2}(a) versus
 (b) and (d) versus (e). Therefore, it needs only
 to consider either $S^{\upa\upa}_{\alf\alf}$ or
 $S^{\dwa\dwa}_{\alf\alf}$.
%%%%%
 For $S^{\upa\upa}_{\alf\alf}$ in \Fig{GRA2}(a)
 and (d), it is found that both the Coulomb
 correlation in the dots and the spin
 polarization of the electrodes will enhance
 the Fano factor. The shot noise is
 of NINT $<$ SDCB $<$ DDCB, consistent with
 the Coulomb correlation strength.
%%%%%%%%%
 The only exception in terms of
 spin polarization is that shown by
 the solid curve in \Fig{GRA2}(a), for the
 weak dot-dot tunnel coupling and in the
 absence of many-body Coulomb interaction.
  In this case the Fano factor does not
 monotonically increase with the
 polarization degree of the electrodes.
%%%%%
 This behavior originates from the same reason
 as that leading to the shoulder behavior of
 the charge current in \Fig{GRA1}(c).

 Remarkably, a profound, strong super-Poisson
 behavior can be developed by increasing
 the dot-dot tunnel coupling for both the
 SDCB and DDCB cases, provided the
 electrodes are properly spin polarized,
 see the dashed and dotted curves in \Fig{GRA2}(d).
%%%%%
 This novel behavior can be understood in terms of
 the so-called {\it dynamical spin blockade} mechanism.
 Take $S^{\upa\upa}_{\alf\alf}$ again for illustration.
 If the electrode is sufficiently spin-up polarized,
 the spin-up electrons can more easily pass through
 the two dots than the spin-down electrons.
 Furthermore, the SDCB (DDCB) does not allow double
 occupation in single dot (double dots). Then a
 mechanism of \emph{fast-to-slow channels} is developed,
 which results in a bunching behavior and also the
 super-Poisson statistics of the transport electrons.

 Note that the strong Coulomb interaction is
 essential to the bunching behavior of tunneling
 events, otherwise the spin-up and spin-down
 electrons will transport independently, and
 will not result in the super-Poisson statistics
 at all \cite{Bla001}.
 Also, a relatively strong dot-dot coupling is
 required for the super-poisson noise in the coupled
 double dots system.
%%%%%%
 If the dot-dot coupling is weak, electron will
 attempt to stay longer in the left dot.
 This will weaken the bunching behavior,
 thus lead all the noise components to sub-Poisson,
 see the dashed and dotted curves in \Fig{GRA2}(a)
 and (b).

 The mechanism of dynamic spin blockade
 plays also an essential role in a
 three-terminal quantum
 dot \cite{Cot04206801,Cot04115315},
 and in an interacting quantum dot with
 intradot spin-flip scattering connected
 to FM electrodes \cite{Dju0571}.
 Meanwhile, a similar mechanism called
 dynamic charge blockade is responsible for
 the super-Poisson noise in interacting
 two-channel systems \cite{Kie0437,Kie03125320,Wan07125416},
 as well as double quantum dots
 structures \cite{Kie07206602,San08035409}.

\subsection{Mutual spin correlation shot noise}

 We now turn to the mutual spin correlation shot
 noise ($S^{\upa\dwa}_{\alf\alf}$ or
 $S^{\upa\dwa}_{\alf\alf}$), which are
 symmetric to the polarization,
 as shown in \Fig{GRA2}(c) and (f).
%%%%%
 For noninteracting dots, the mutual
 correlation between opposite spin currents are zero,
 as shown by the solid curves in \Fig{GRA2}(c) and (f).
%%%%%
 This is simply because the spin-up and spin-down
 currents are uncorrelated in the absence
 of Coulomb interaction, despite the electrodes are
 spin polarized.
%%%%%
 Similar behavior was also found
 in \cite{Sau04106601}, where the mutual
 correlation is zero for noninteracting single dot
 connected with normal electrodes.

 For interacting dots, i.e., the SDCB and DDCB
 shown in \Fig{GRA2} (c) and (f), the mutual
 correlation of spin-up and spin-down electrons
 reveals relatively complex features which can
 be either \emph{positive} or \emph{negative}.
 The main features are as follows. (i) For SDCB,
 negative mutual-correlation is found for the
 weakly coupled dots, if the electrodes are
 weakly or moderately polarized; but positive
 correlation can be formed if the electrodes
 are strongly polarized.
 (ii) Also for SDCB, the mutual correlation is
 \emph{fully negative} for strongly coupled dots,
 independent of the polarization degree of the
 electrodes.
 (iii) For DDCB, in the case of weak dot-dot
 coupling, the mutual correlation is fully
 negative.
 (iv) Again for DDCB, but in the case of strong
 dot-dot coupling, the mutual correlation is
 positive and can be of strong super-Poisson
 for electrodes sufficiently polarized, while
 it is negative for electrodes weakly or
 non-polarized.

 In this context, we notice that the
 super-Poissonian self-correlation does not necessarily
 imply a positive mutual correlation, as shown by
 the dashed curve in \Fig{GRA2}(f) for the SDCB,
 where the mutual correlation, despite the
 polarization degree of the electrodes, is fully
 negative.
 A simple and unambiguous understanding for the
 above positive and negative mutual correlation
 is yet to be found. It seems more subtle than
 the super-Poisson statistics of the
 self-correlation of current.

 \subsection{\bf Total charge and spin current noises}

\begin{figure}
 \includegraphics*[scale=0.9]{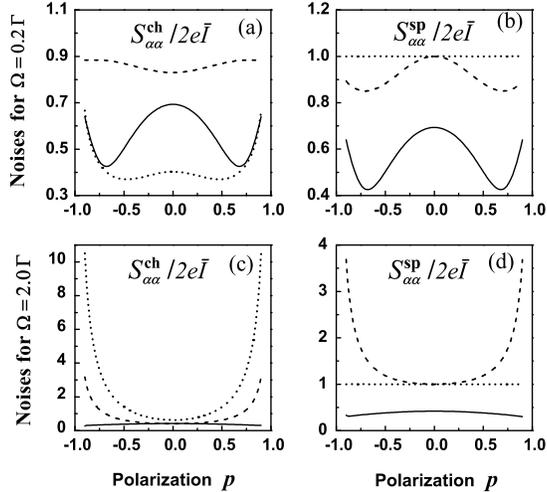}
 \caption{\label{GRA3}Total charge and
 spin current noises, obtained by
 appropriate combination of the individual
 noise components as shown in \Fig{GRA2}
 according to \Eq{Stot}. The plot parameters
 are the same as in \Fig{GRA2}.}
 \end{figure}

 The total charge and spin current noises
 are nothing but the combination of those
 components of the spin-dependent noises,
 according to \Eq{Stot}.
 The results are displayed in \Fig{GRA3}, where
 most features such as the sub-Poisson and
 super-Poisson behaviors can be accordingly
 understood in terms of the above interpretation
 to the partial noises.

 A noticeable result is the spin current noise
 resulting from the DDCB. We find that it is
 constantly Poissonian, regardless the polarization
 degree and dot-dot coupling, as shown by the
 dotted curves in \Fig{GRA3}(b) and (d).
%%%%%%
 Similar Poissonian spin current noise was also
 found in \cite{Sau04106601},
 for single quantum dot in the Coulomb blockade
 regime.

 \begin{table*}
 \caption{\label{table2}Spin-dependent noise,
 as well as the total charge and spin current
 noises for $p=0$ in different cases.}
 \begin{tabular}{ccccc}\hline\hline
 Noises
 & $S_{\alf\alf}^{\upa\upa}$ ($S_{\alf\alf}^{\dwa\dwa}$)
 & $S_{\alf\alf}^{\upa\dwa}$ ($S_{\alf\alf}^{\dwa\upa}$)
 & $S_{\alf\alf}^{\rm ch}$
 & $S_{\alf\alf}^{\rm sp}$
 \\  \hline
 case (i)
 &
 $e\bar{I}_0\frac{\Gam^4-2\Gam^2\Omg^2+8\Omg^4}
 {(\Gam^2+4\Omg^2)^2}$
 &
 0
 &
 $2e\bar{I}_0\frac{\Gam^4-2\Gam^2\Omg^2+8\Omg^4}
 {(\Gam^2+4\Omg^2)^2}$
 &
 $2e\bar{I}_0\frac{\Gam^4-2\Gam^2\Omg^2+8\Omg^4}
 {(\Gam^2+4\Omg^2)^2}$
 \vspace{1ex} \\
 case (ii)
 &
 $e\bar{I}_\rms\frac{9\Gam^4
 +14\Gam^2\Omg^2+28\Omg^4}
 {9(\Gam^2+2\Omg^2)^2}$
 &
 $-2e\bar{I}_\rms\frac{(11\Gam^2+4\Omg^2)\Omg^2}
 {9(\Gam^2+2\Omg^2)^2}$
 &
 $2e\bar{I}_\rms\frac{9\Gam^4-8\Gam^2\Omg^2+20\Omg^4}
 {9(\Gam^2+2\Omg^2)^2}$
 &
 $2e\bar{I}_\rms$
 \vspace{1ex} \\
 case (iii)
 &
 $e\bar{I}_\rmd\frac{\Gam^4+6\Gam^2\Omg^2+84\Omg^4}
 {(\Gam^2+10\Omg^2)^2}$
 &
 $-2e\bar{I}_\rmd\frac{(7\Gam^2+8\Omg^2)\Omg^2}
 {(\Gam^2+10\Omg^2)^2}$
 &
 $2e\bar{I}_\rmd\frac{\Gam^4-8\Gam^2\Omg^2+68\Omg^4}
 {(\Gam^2+10\Omg^2)^2}$
 &
 $2e\bar{I}_\rmd$ \vspace{1ex} \\  \hline\hline
 \end{tabular}
 \end{table*}

 To complete this section, we consider all
 kinds of noises discussed above in the limit
 of $p=0$. In this unpolarized situation,
 simple analytic results can be obtained,
 as shown in table \ref{table2}, which
 serves as a complement to the numerical
 results in \Fig{GRA2} and \Fig{GRA3}.
%%%%%
 The total charge current noise in the NINT case
 differs from the ones in \cite{Ela02289,%
 Sun9910748,Kie06033312} by an overall factor of 2
 (implied in $\bar{I}_0$, see table \ref{table}).
 This factor originates from the spin degree of freedom,
 which was ignored there.
%%%%%
 For interacting cases, under symmetric coupling
 ($\GamL=\GamR$) and resonant levels
 ($E_{\rm L}=E_{\rm R}$), the total charge
 current noises are found here to be
 sub-Poissonian, in either the SDCB or the
 DDCB cases.
%%%%%
 Very interestingly, however, if $E_{\rm L}\neq E_{\rm R}$
 and $\GamL>\GamR$, the interplay of quantum
 coherence and the DDCB would develop a dynamical
 charge blockade mechanism, which then result
 in a remarkable super-Poissonian noise.
%%%%%
 Our result in the DDCB case slightly differs
 from the ones in \cite{Kie07206602,Ela02289}.
 Again, the difference arises from the spin
 degree of freedom, which was neglected in the
 previous work.
%%%%%

 \section{\label{thsec6}Conclusion }

 To summarize, based on an efficient particle-number
 resolved quantum master equation with
 inclusion of the spin degrees of freedom,
 we investigated the spin-dependent noises in transport
 through a pair of coupled quantum dots
 connected with ferromagnetic electrodes.
%%%%%
 The modulation of the dot-dot coupling,
 and the interplay between Coulomb
 interactions and spin polarization in the
 electrodes give rise to rich noise behaviors,
 such as super-Poissonian, constant Poissonian
 noises, as well as positive and negative
 mutual correlations.
 These unique noise features can serve as additional tool
 in experiments for revealing the intrinsic
 dot-dot coupling, as well as the
 Coulomb interactions that involved.
 The results presented in this work were carried out at zero
 temperature. At finite temperatures, backward processes are
 possible, which will in general reduce the current,
 but enhance the noise, as clearly demonstrated
 in \cite{Wan07125416}.

 Finally, we briefly discuss the measurement of
 spin-dependent noise.
%%%%%
 It was shown in \cite{For05016601}
 that in hybrid ferromagnetic-normal metal
 structures, the spin current noise exerts a
 fluctuating spin torque on the magnetization
 vector, which causes an observable magnetization noise.
%%%%%
 By measuring this magnetization noise, which
 has been realized experimentally \cite{Cov04184406},
 one then obtains the spin current noise.

\begin{acknowledgments}
 This work was supported by the RGC (604007) of
 Hong Kong, the National Natural Science
 Foundation of China under grants No.\ 60425412
 and No.\ 90503013, and the Major State Basic
 Research Project under grant No.2006CB921201.
\end{acknowledgments}

%\bibliographystyle{/home/yan/refs/aip}
%\bibliography{/home/yan/refs/bibrefs}
%\bibliography{E:/bibliography/bibrefs}
%\bibliographystyle{unsrt}
%\bibliography{/disk3/yan/refs/bibrefs}
%\input{mainbbl}

%\bibliographystyle{IOP}
%\bibliography{E:/bibliography/bibrefs}

\begin{thebibliography}{47}
\expandafter\ifx\csname
natexlab\endcsname\relax\def\natexlab#1{#1}\fi
\expandafter\ifx\csname bibnamefont\endcsname\relax
  \def\bibnamefont#1{#1}\fi
\expandafter\ifx\csname bibfnamefont\endcsname\relax
  \def\bibfnamefont#1{#1}\fi
\expandafter\ifx\csname citenamefont\endcsname\relax
  \def\citenamefont#1{#1}\fi
\expandafter\ifx\csname url\endcsname\relax
  \def\url#1{\texttt{#1}}\fi
\expandafter\ifx\csname urlprefix\endcsname\relax\def\urlprefix{URL
}\fi \providecommand{\bibinfo}[2]{#2}
\providecommand{\eprint}[2][]{\url{#2}}

\bibitem[{\citenamefont{Blanter and {B\"{u}ttiker}}(2000)}]{Bla001}
\bibinfo{author}{\bibfnamefont{Y.~M.} \bibnamefont{Blanter}} \bibnamefont{and}
  \bibinfo{author}{\bibfnamefont{M.}~\bibnamefont{{B\"{u}ttiker}}},
  \bibinfo{journal}{Phys. Rep.} \textbf{\bibinfo{volume}{336}},
  \bibinfo{pages}{1} (\bibinfo{year}{2000}).

\bibitem[{\citenamefont{Nazarov}(2003)}]{Naz03}
\bibinfo{author}{\bibfnamefont{Y.~V.} \bibnamefont{Nazarov}},
  \emph{\bibinfo{title}{Quantum Noise in Mesoscopic Physics}}
  (\bibinfo{publisher}{Kluwer}, \bibinfo{address}{Dordrecht},
  \bibinfo{year}{2003}).

\bibitem[{\citenamefont{Blencowe}(2004)}]{Ble04159}
\bibinfo{author}{\bibfnamefont{M.}~\bibnamefont{Blencowe}},
  \bibinfo{journal}{Phys. Rep.} \textbf{\bibinfo{volume}{395}},
  \bibinfo{pages}{159} (\bibinfo{year}{2004}).

\bibitem[{\citenamefont{LaHaye et~al.}(2004)\citenamefont{LaHaye, Buu,
  Camarota, and Schwab}}]{LaH0474}
\bibinfo{author}{\bibfnamefont{M.~D.} \bibnamefont{LaHaye}},
  \bibinfo{author}{\bibfnamefont{O.}~\bibnamefont{Buu}},
  \bibinfo{author}{\bibfnamefont{B.}~\bibnamefont{Camarota}}, \bibnamefont{and}
  \bibinfo{author}{\bibfnamefont{K.~C.} \bibnamefont{Schwab}},
  \bibinfo{journal}{Science} \textbf{\bibinfo{volume}{304}},
  \bibinfo{pages}{74} (\bibinfo{year}{2004}).

\bibitem[{\citenamefont{{Novotn\'{y}} et~al.}(2004)\citenamefont{{Novotn\'{y}},
  Donarini, Flindt, and Jauho}}]{Nov04248302}
\bibinfo{author}{\bibfnamefont{T.}~\bibnamefont{{Novotn\'{y}}}},
  \bibinfo{author}{\bibfnamefont{A.}~\bibnamefont{Donarini}},
  \bibinfo{author}{\bibfnamefont{C.}~\bibnamefont{Flindt}}, \bibnamefont{and}
  \bibinfo{author}{\bibfnamefont{A.-P.} \bibnamefont{Jauho}},
  \bibinfo{journal}{Phys. Rev. Lett.} \textbf{\bibinfo{volume}{92}},
  \bibinfo{pages}{248302} (\bibinfo{year}{2004}).

\bibitem[{\citenamefont{Flindt et~al.}(2005)\citenamefont{Flindt,
  {Novotn\'{y}}, and Jauho}}]{Fli05411}
\bibinfo{author}{\bibfnamefont{C.}~\bibnamefont{Flindt}},
  \bibinfo{author}{\bibfnamefont{T.}~\bibnamefont{{Novotn\'{y}}}},
  \bibnamefont{and} \bibinfo{author}{\bibfnamefont{A.-P.} \bibnamefont{Jauho}},
  \bibinfo{journal}{Physica E} \textbf{\bibinfo{volume}{29}},
  \bibinfo{pages}{411} (\bibinfo{year}{2005}).

\bibitem[{\citenamefont{Braun et~al.}(2006)\citenamefont{Braun, {K\"{o}nig},
  and Martinek}}]{Bra0601366}
\bibinfo{author}{\bibfnamefont{M.}~\bibnamefont{Braun}},
  \bibinfo{author}{\bibfnamefont{J.}~\bibnamefont{{K\"{o}nig}}},
  \bibnamefont{and} \bibinfo{author}{\bibfnamefont{J.}~\bibnamefont{Martinek}}
  (\bibinfo{year}{2006}), \bibinfo{note}{cond-mat/0601366}.

\bibitem[{\citenamefont{Gurvitz et~al.}(2005)\citenamefont{Gurvitz, Mozyrsky,
  and Berman}}]{Gur05205341}
\bibinfo{author}{\bibfnamefont{S.~A.} \bibnamefont{Gurvitz}},
  \bibinfo{author}{\bibfnamefont{D.}~\bibnamefont{Mozyrsky}}, \bibnamefont{and}
  \bibinfo{author}{\bibfnamefont{G.~P.} \bibnamefont{Berman}},
  \bibinfo{journal}{Phys. Rev. B} \textbf{\bibinfo{volume}{72}},
  \bibinfo{pages}{205341} (\bibinfo{year}{2005}).

\bibitem[{\citenamefont{Belzig and Zareyan}(2004)}]{Bel04140407}
\bibinfo{author}{\bibfnamefont{W.}~\bibnamefont{Belzig}} \bibnamefont{and}
  \bibinfo{author}{\bibfnamefont{M.}~\bibnamefont{Zareyan}},
  \bibinfo{journal}{Phys. Rev. B} \textbf{\bibinfo{volume}{69}},
  \bibinfo{pages}{140407(R)} (\bibinfo{year}{2004}).

\bibitem[{\citenamefont{Elattari and Gurvitz}(2002)}]{Ela02289}
\bibinfo{author}{\bibfnamefont{B.}~\bibnamefont{Elattari}} \bibnamefont{and}
  \bibinfo{author}{\bibfnamefont{S.~A.} \bibnamefont{Gurvitz}},
  \bibinfo{journal}{Phys. Lett. A} \textbf{\bibinfo{volume}{292}},
  \bibinfo{pages}{289} (\bibinfo{year}{2002}).

\bibitem[{\citenamefont{Aghassi et~al.}(2006)\citenamefont{Aghassi, Thielmann,
  Hettler, and {Sch\"{o}n}}}]{Agh06195323}
\bibinfo{author}{\bibfnamefont{J.}~\bibnamefont{Aghassi}},
  \bibinfo{author}{\bibfnamefont{A.}~\bibnamefont{Thielmann}},
  \bibinfo{author}{\bibfnamefont{M.~H.} \bibnamefont{Hettler}},
  \bibnamefont{and}
  \bibinfo{author}{\bibfnamefont{G.}~\bibnamefont{{Sch\"{o}n}}},
  \bibinfo{journal}{Phys. Rev. B} \textbf{\bibinfo{volume}{73}},
  \bibinfo{pages}{195323} (\bibinfo{year}{2006}).

\bibitem[{\citenamefont{Sun and Milburn}(1999)}]{Sun9910748}
\bibinfo{author}{\bibfnamefont{H.~B.} \bibnamefont{Sun}} \bibnamefont{and}
  \bibinfo{author}{\bibfnamefont{G.~J.} \bibnamefont{Milburn}},
  \bibinfo{journal}{Phys. Rev. B} \textbf{\bibinfo{volume}{59}},
  \bibinfo{pages}{10748} (\bibinfo{year}{1999}).

\bibitem[{\citenamefont{Aguado and Brandes}(2004)}]{Agu04206601}
\bibinfo{author}{\bibfnamefont{R.}~\bibnamefont{Aguado}} \bibnamefont{and}
  \bibinfo{author}{\bibfnamefont{T.}~\bibnamefont{Brandes}},
  \bibinfo{journal}{Phys. Rev. Lett.} \textbf{\bibinfo{volume}{92}},
  \bibinfo{pages}{206601} (\bibinfo{year}{2004}).

\bibitem[{\citenamefont{Makhlin et~al.}(2001)\citenamefont{Makhlin,
  {Sch\"{o}n}, and Shnirman}}]{Mak01357}
\bibinfo{author}{\bibfnamefont{Y.}~\bibnamefont{Makhlin}},
  \bibinfo{author}{\bibfnamefont{G.}~\bibnamefont{{Sch\"{o}n}}},
  \bibnamefont{and} \bibinfo{author}{\bibfnamefont{A.}~\bibnamefont{Shnirman}},
  \bibinfo{journal}{Rev. Mod. Phys.} \textbf{\bibinfo{volume}{73}},
  \bibinfo{pages}{357} (\bibinfo{year}{2001}).

\bibitem[{\citenamefont{Li et~al.}(2005{\natexlab{a}})\citenamefont{Li, Cui,
  and Yan}}]{Li05066803}
\bibinfo{author}{\bibfnamefont{X.~Q.} \bibnamefont{Li}},
  \bibinfo{author}{\bibfnamefont{P.}~\bibnamefont{Cui}}, \bibnamefont{and}
  \bibinfo{author}{\bibfnamefont{Y.~J.} \bibnamefont{Yan}},
  \bibinfo{journal}{Phys. Rev. Lett.} \textbf{\bibinfo{volume}{94}},
  \bibinfo{pages}{066803} (\bibinfo{year}{2005}{\natexlab{a}}).

\bibitem[{\citenamefont{Gurvitz et~al.}(2003)\citenamefont{Gurvitz, Fedichkin,
  Mozyrsky, and Berman}}]{Gur03066801}
\bibinfo{author}{\bibfnamefont{S.~A.} \bibnamefont{Gurvitz}},
  \bibinfo{author}{\bibfnamefont{L.}~\bibnamefont{Fedichkin}},
  \bibinfo{author}{\bibfnamefont{D.}~\bibnamefont{Mozyrsky}}, \bibnamefont{and}
  \bibinfo{author}{\bibfnamefont{G.~P.} \bibnamefont{Berman}},
  \bibinfo{journal}{Phys. Rev. Lett.} \textbf{\bibinfo{volume}{91}},
  \bibinfo{pages}{066801} (\bibinfo{year}{2003}).

\bibitem[{\citenamefont{Stace and Barrett}(2004)}]{Sta04136802}
\bibinfo{author}{\bibfnamefont{T.~M.} \bibnamefont{Stace}} \bibnamefont{and}
  \bibinfo{author}{\bibfnamefont{S.~D.} \bibnamefont{Barrett}},
  \bibinfo{journal}{Phys. Rev. Lett.} \textbf{\bibinfo{volume}{92}},
  \bibinfo{pages}{136802} (\bibinfo{year}{2004}).

\bibitem[{\citenamefont{Wang et~al.}(2004)\citenamefont{Wang, Wang, and
  Guo}}]{Wan04153301}
\bibinfo{author}{\bibfnamefont{B.}~\bibnamefont{Wang}},
  \bibinfo{author}{\bibfnamefont{J.}~\bibnamefont{Wang}}, \bibnamefont{and}
  \bibinfo{author}{\bibfnamefont{H.}~\bibnamefont{Guo}},
  \bibinfo{journal}{Phys. Rev. B} \textbf{\bibinfo{volume}{69}},
  \bibinfo{pages}{153301} (\bibinfo{year}{2004}).

\bibitem[{\citenamefont{Dong et~al.}(2005)\citenamefont{Dong, Cui, and
  Lei}}]{Don05066601}
\bibinfo{author}{\bibfnamefont{B.}~\bibnamefont{Dong}},
  \bibinfo{author}{\bibfnamefont{H.~L.} \bibnamefont{Cui}}, \bibnamefont{and}
  \bibinfo{author}{\bibfnamefont{X.~L.} \bibnamefont{Lei}},
  \bibinfo{journal}{Phys. Rev. Lett.} \textbf{\bibinfo{volume}{94}},
  \bibinfo{pages}{066601} (\bibinfo{year}{2005}).

\bibitem[{\citenamefont{Djuric and Search}(2006)}]{Dju06115327}
\bibinfo{author}{\bibfnamefont{I.}~\bibnamefont{Djuric}} \bibnamefont{and}
  \bibinfo{author}{\bibfnamefont{C.~P.} \bibnamefont{Search}},
  \bibinfo{journal}{Phys. Rev. B} \textbf{\bibinfo{volume}{74}},
  \bibinfo{pages}{115327} (\bibinfo{year}{2006}).

\bibitem[{\citenamefont{Sauret and Feinberg}(2004)}]{Sau04106601}
\bibinfo{author}{\bibfnamefont{O.}~\bibnamefont{Sauret}} \bibnamefont{and}
  \bibinfo{author}{\bibfnamefont{D.}~\bibnamefont{Feinberg}},
  \bibinfo{journal}{Phys. Rev. Lett.} \textbf{\bibinfo{volume}{92}},
  \bibinfo{pages}{106601} (\bibinfo{year}{2004}).

\bibitem[{\citenamefont{Li et~al.}(2005{\natexlab{b}})\citenamefont{Li, Luo,
  Yang, Cui, and Yan}}]{Li05205304}
\bibinfo{author}{\bibfnamefont{X.~Q.} \bibnamefont{Li}},
  \bibinfo{author}{\bibfnamefont{J.~Y.} \bibnamefont{Luo}},
  \bibinfo{author}{\bibfnamefont{Y.~G.} \bibnamefont{Yang}},
  \bibinfo{author}{\bibfnamefont{P.}~\bibnamefont{Cui}}, \bibnamefont{and}
  \bibinfo{author}{\bibfnamefont{Y.~J.} \bibnamefont{Yan}},
  \bibinfo{journal}{Phys. Rev. B} \textbf{\bibinfo{volume}{71}},
  \bibinfo{pages}{205304} (\bibinfo{year}{2005}{\natexlab{b}}).

\bibitem[{\citenamefont{Luo et~al.}(2007)\citenamefont{Luo, Li, and
  Yan}}]{Luo07085325}
\bibinfo{author}{\bibfnamefont{J.~Y.} \bibnamefont{Luo}},
  \bibinfo{author}{\bibfnamefont{X.-Q.} \bibnamefont{Li}}, \bibnamefont{and}
  \bibinfo{author}{\bibfnamefont{Y.~J.} \bibnamefont{Yan}},
  \bibinfo{journal}{Phys. Rev. B} \textbf{\bibinfo{volume}{76}},
  \bibinfo{pages}{085325} (\bibinfo{year}{2007}).

\bibitem[{\citenamefont{Gurvitz and Prager}(1996)}]{Gur9615932}
\bibinfo{author}{\bibfnamefont{S.~A.} \bibnamefont{Gurvitz}} \bibnamefont{and}
  \bibinfo{author}{\bibfnamefont{Y.~S.} \bibnamefont{Prager}},
  \bibinfo{journal}{Phys. Rev. B} \textbf{\bibinfo{volume}{53}},
  \bibinfo{pages}{15932} (\bibinfo{year}{1996}).

\bibitem[{\citenamefont{Ono et~al.}(2002)\citenamefont{Ono, Austing, Tokura,
  and Tarucha}}]{Ono021313}
\bibinfo{author}{\bibfnamefont{K.}~\bibnamefont{Ono}},
  \bibinfo{author}{\bibfnamefont{D.~G.} \bibnamefont{Austing}},
  \bibinfo{author}{\bibfnamefont{Y.}~\bibnamefont{Tokura}}, \bibnamefont{and}
  \bibinfo{author}{\bibfnamefont{S.}~\bibnamefont{Tarucha}},
  \bibinfo{journal}{Science} \textbf{\bibinfo{volume}{297}},
  \bibinfo{pages}{1313} (\bibinfo{year}{2002}).

\bibitem[{\citenamefont{McClure et~al.}(2007)\citenamefont{McClure, DiCarlo,
  Zhang, Engel, Marcus, Hanson, and Gossard}}]{McC07056801}
\bibinfo{author}{\bibfnamefont{D.~T.} \bibnamefont{McClure}},
  \bibinfo{author}{\bibfnamefont{L.}~\bibnamefont{DiCarlo}},
  \bibinfo{author}{\bibfnamefont{Y.}~\bibnamefont{Zhang}},
  \bibinfo{author}{\bibfnamefont{H.-A.} \bibnamefont{Engel}},
  \bibinfo{author}{\bibfnamefont{C.~M.} \bibnamefont{Marcus}},
  \bibinfo{author}{\bibfnamefont{M.~P.} \bibnamefont{Hanson}},
  \bibnamefont{and} \bibinfo{author}{\bibfnamefont{A.~C.}
  \bibnamefont{Gossard}}, \bibinfo{journal}{Phys. Rev. Lett.}
  \textbf{\bibinfo{volume}{98}}, \bibinfo{pages}{056801}
  (\bibinfo{year}{2007}).

\bibitem[{\citenamefont{Yan}(1998)}]{Yan982721}
\bibinfo{author}{\bibfnamefont{Y.~J.} \bibnamefont{Yan}},
  \bibinfo{journal}{Phys. Rev. A} \textbf{\bibinfo{volume}{58}},
  \bibinfo{pages}{2721} (\bibinfo{year}{1998}).

\bibitem[{\citenamefont{MacDonald}(1962)}]{Mac62}
\bibinfo{author}{\bibfnamefont{D.~K.~C.} \bibnamefont{MacDonald}},
  \emph{\bibinfo{title}{Noise and Fluctuations: An Introduction}}
  (\bibinfo{publisher}{Wiley}, \bibinfo{address}{New York},
  \bibinfo{year}{1962}), \bibinfo{note}{ch.\ 2.2.1}.

\bibitem[{\citenamefont{Mozyrsky et~al.}(2002)\citenamefont{Mozyrsky,
  Fedichkin, Gurvitz, and Berman}}]{Moz02161313}
\bibinfo{author}{\bibfnamefont{D.}~\bibnamefont{Mozyrsky}},
  \bibinfo{author}{\bibfnamefont{L.}~\bibnamefont{Fedichkin}},
  \bibinfo{author}{\bibfnamefont{S.~A.} \bibnamefont{Gurvitz}},
  \bibnamefont{and} \bibinfo{author}{\bibfnamefont{G.~P.}
  \bibnamefont{Berman}}, \bibinfo{journal}{Phys. Rev. B}
  \textbf{\bibinfo{volume}{66}}, \bibinfo{pages}{161313}
  (\bibinfo{year}{2002}).

\bibitem[{\citenamefont{Wunsch et~al.}(2005)\citenamefont{Wunsch, Braun,
  {K\"{o}nig}, and Pfannkuche}}]{Wun05205319}
\bibinfo{author}{\bibfnamefont{B.}~\bibnamefont{Wunsch}},
  \bibinfo{author}{\bibfnamefont{M.}~\bibnamefont{Braun}},
  \bibinfo{author}{\bibfnamefont{J.}~\bibnamefont{{K\"{o}nig}}},
  \bibnamefont{and}
  \bibinfo{author}{\bibfnamefont{D.}~\bibnamefont{Pfannkuche}},
  \bibinfo{journal}{Phys. Rev. B} \textbf{\bibinfo{volume}{72}},
  \bibinfo{pages}{205319} (\bibinfo{year}{2005}).

\bibitem[{\citenamefont{Cui et~al.}(2006)\citenamefont{Cui, Li, Shao, and
  Yan}}]{Cui06449}
\bibinfo{author}{\bibfnamefont{P.}~\bibnamefont{Cui}},
  \bibinfo{author}{\bibfnamefont{X.~Q.} \bibnamefont{Li}},
  \bibinfo{author}{\bibfnamefont{J.~S.} \bibnamefont{Shao}}, \bibnamefont{and}
  \bibinfo{author}{\bibfnamefont{Y.~J.} \bibnamefont{Yan}},
  \bibinfo{journal}{Phys. Lett. A} \textbf{\bibinfo{volume}{357}},
  \bibinfo{pages}{449} (\bibinfo{year}{2006}).

\bibitem[{\citenamefont{Jin et~al.}(2007)\citenamefont{Jin, Welack, Luo, Li,
  Cui, Xu, and Yan}}]{Jin07134113}
\bibinfo{author}{\bibfnamefont{J.~S.} \bibnamefont{Jin}},
  \bibinfo{author}{\bibfnamefont{S.}~\bibnamefont{Welack}},
  \bibinfo{author}{\bibfnamefont{J.~Y.} \bibnamefont{Luo}},
  \bibinfo{author}{\bibfnamefont{X.~Q.} \bibnamefont{Li}},
  \bibinfo{author}{\bibfnamefont{P.}~\bibnamefont{Cui}},
  \bibinfo{author}{\bibfnamefont{R.~X.} \bibnamefont{Xu}}, \bibnamefont{and}
  \bibinfo{author}{\bibfnamefont{Y.~J.} \bibnamefont{Yan}},
  \bibinfo{journal}{J. Chem. Phys.} \textbf{\bibinfo{volume}{126}},
  \bibinfo{pages}{134113} (\bibinfo{year}{2007}).

\bibitem[{\citenamefont{Petta et~al.}(2005)\citenamefont{Petta, Johnson,
  Taylor, Laird, Yacoby, Lukin, Marcus, Hanson, and Gossard}}]{Pet052180}
\bibinfo{author}{\bibfnamefont{J.~R.} \bibnamefont{Petta}},
  \bibinfo{author}{\bibfnamefont{A.~C.} \bibnamefont{Johnson}},
  \bibinfo{author}{\bibfnamefont{J.~M.} \bibnamefont{Taylor}},
  \bibinfo{author}{\bibfnamefont{E.~A.} \bibnamefont{Laird}},
  \bibinfo{author}{\bibfnamefont{A.}~\bibnamefont{Yacoby}},
  \bibinfo{author}{\bibfnamefont{M.~D.} \bibnamefont{Lukin}},
  \bibinfo{author}{\bibfnamefont{C.~M.} \bibnamefont{Marcus}},
  \bibinfo{author}{\bibfnamefont{M.~P.} \bibnamefont{Hanson}},
  \bibnamefont{and} \bibinfo{author}{\bibfnamefont{A.~C.}
  \bibnamefont{Gossard}}, \bibinfo{journal}{Science}
  \textbf{\bibinfo{volume}{309}}, \bibinfo{pages}{2180} (\bibinfo{year}{2005}).

\bibitem[{\citenamefont{Yan and Xu}(2005)}]{Yan05187}
\bibinfo{author}{\bibfnamefont{Y.~J.} \bibnamefont{Yan}} \bibnamefont{and}
  \bibinfo{author}{\bibfnamefont{R.~X.} \bibnamefont{Xu}},
  \bibinfo{journal}{Annu. Rev. Phys. Chem.} \textbf{\bibinfo{volume}{56}},
  \bibinfo{pages}{187} (\bibinfo{year}{2005}).

\bibitem[{\citenamefont{Xu and Yan}(2002)}]{Xu029196}
\bibinfo{author}{\bibfnamefont{R.~X.} \bibnamefont{Xu}} \bibnamefont{and}
  \bibinfo{author}{\bibfnamefont{Y.~J.} \bibnamefont{Yan}},
  \bibinfo{journal}{J. Chem. Phys.} \textbf{\bibinfo{volume}{116}},
  \bibinfo{pages}{9196} (\bibinfo{year}{2002}).

\bibitem[{\citenamefont{Bagrets and Nazarov}(2003)}]{Bag03085316}
\bibinfo{author}{\bibfnamefont{D.~A.} \bibnamefont{Bagrets}} \bibnamefont{and}
  \bibinfo{author}{\bibfnamefont{Y.~V.} \bibnamefont{Nazarov}},
  \bibinfo{journal}{Phys. Rev. B} \textbf{\bibinfo{volume}{67}},
  \bibinfo{pages}{085316} (\bibinfo{year}{2003}).

\bibitem[{\citenamefont{Cottet et~al.}(2004{\natexlab{a}})\citenamefont{Cottet,
  Belzig, and Bruder}}]{Cot04115315}
\bibinfo{author}{\bibfnamefont{A.}~\bibnamefont{Cottet}},
  \bibinfo{author}{\bibfnamefont{W.}~\bibnamefont{Belzig}}, \bibnamefont{and}
  \bibinfo{author}{\bibfnamefont{C.}~\bibnamefont{Bruder}},
  \bibinfo{journal}{Phys. Rev. B} \textbf{\bibinfo{volume}{70}},
  \bibinfo{pages}{115315} (\bibinfo{year}{2004}{\natexlab{a}}).

\bibitem[{\citenamefont{Cottet et~al.}(2004{\natexlab{b}})\citenamefont{Cottet,
  Belzig, and Bruder}}]{Cot04206801}
\bibinfo{author}{\bibfnamefont{A.}~\bibnamefont{Cottet}},
  \bibinfo{author}{\bibfnamefont{W.}~\bibnamefont{Belzig}}, \bibnamefont{and}
  \bibinfo{author}{\bibfnamefont{C.}~\bibnamefont{Bruder}},
  \bibinfo{journal}{Phys. Rev. Lett.} \textbf{\bibinfo{volume}{92}},
  \bibinfo{pages}{206801} (\bibinfo{year}{2004}{\natexlab{b}}).

\bibitem[{\citenamefont{Djuric et~al.}(2005)\citenamefont{Djuric, Dong, and
  Cui}}]{Dju0571}
\bibinfo{author}{\bibfnamefont{I.}~\bibnamefont{Djuric}},
  \bibinfo{author}{\bibfnamefont{B.}~\bibnamefont{Dong}}, \bibnamefont{and}
  \bibinfo{author}{\bibfnamefont{H.-L.} \bibnamefont{Cui}},
  \bibinfo{journal}{IEEE Transactions on Nanotechnology}
  \textbf{\bibinfo{volume}{4}}, \bibinfo{pages}{71} (\bibinfo{year}{2005}).

\bibitem[{\citenamefont{{Kie{\ss}lich}
  et~al.}(2004)\citenamefont{{Kie{\ss}lich}, Sprekeler, and
  {Sch\"{o}ll}}}]{Kie0437}
\bibinfo{author}{\bibfnamefont{G.}~\bibnamefont{{Kie{\ss}lich}}},
  \bibinfo{author}{\bibfnamefont{H.}~\bibnamefont{Sprekeler}},
  \bibnamefont{and}
  \bibinfo{author}{\bibfnamefont{E.}~\bibnamefont{{Sch\"{o}ll}}},
  \bibinfo{journal}{Semicond. Sci. Technol.} \textbf{\bibinfo{volume}{19}},
  \bibinfo{pages}{S37} (\bibinfo{year}{2004}).

\bibitem[{\citenamefont{{Kie{\ss}lich}
  et~al.}(2003)\citenamefont{{Kie{\ss}lich}, Wacker, and
  {Sch\"{o}ll}}}]{Kie03125320}
\bibinfo{author}{\bibfnamefont{G.}~\bibnamefont{{Kie{\ss}lich}}},
  \bibinfo{author}{\bibfnamefont{A.}~\bibnamefont{Wacker}}, \bibnamefont{and}
  \bibinfo{author}{\bibfnamefont{E.}~\bibnamefont{{Sch\"{o}ll}}},
  \bibinfo{journal}{Phys. Rev. B} \textbf{\bibinfo{volume}{68}},
  \bibinfo{pages}{125320} (\bibinfo{year}{2003}).

\bibitem[{\citenamefont{{Kie{\ss}lich}
  et~al.}(2007)\citenamefont{{Kie{\ss}lich}, {Sch\"{o}ll}, Brandes, Hohls, and
  Haug}}]{Kie07206602}
\bibinfo{author}{\bibfnamefont{G.}~\bibnamefont{{Kie{\ss}lich}}},
  \bibinfo{author}{\bibfnamefont{E.}~\bibnamefont{{Sch\"{o}ll}}},
  \bibinfo{author}{\bibfnamefont{T.}~\bibnamefont{Brandes}},
  \bibinfo{author}{\bibfnamefont{F.}~\bibnamefont{Hohls}}, \bibnamefont{and}
  \bibinfo{author}{\bibfnamefont{R.~J.} \bibnamefont{Haug}},
  \bibinfo{journal}{Phys. Rev. Lett.} \textbf{\bibinfo{volume}{99}},
  \bibinfo{pages}{206602} (\bibinfo{year}{2007}).

\bibitem[{\citenamefont{{S\'{a}nchez} et~al.}(2008)\citenamefont{{S\'{a}nchez},
  Kohler, {H\"{a}nggi}, and Platero}}]{San08035409}
\bibinfo{author}{\bibfnamefont{R.}~\bibnamefont{{S\'{a}nchez}}},
  \bibinfo{author}{\bibfnamefont{S.}~\bibnamefont{Kohler}},
  \bibinfo{author}{\bibfnamefont{P.}~\bibnamefont{{H\"{a}nggi}}},
  \bibnamefont{and} \bibinfo{author}{\bibfnamefont{G.}~\bibnamefont{Platero}},
  \bibinfo{journal}{Phys. Rev. B} \textbf{\bibinfo{volume}{77}},
  \bibinfo{pages}{035409} (\bibinfo{year}{2008}).

\bibitem[{\citenamefont{{Kie{\ss}lich}
  et~al.}(2006)\citenamefont{{Kie{\ss}lich}, Samuelsson, Wacker, and
  {Sch\"{o}ll}}}]{Kie06033312}
\bibinfo{author}{\bibfnamefont{G.}~\bibnamefont{{Kie{\ss}lich}}},
  \bibinfo{author}{\bibfnamefont{P.}~\bibnamefont{Samuelsson}},
  \bibinfo{author}{\bibfnamefont{A.}~\bibnamefont{Wacker}}, \bibnamefont{and}
  \bibinfo{author}{\bibfnamefont{E.}~\bibnamefont{{Sch\"{o}ll}}},
  \bibinfo{journal}{Phys. Rev. B} \textbf{\bibinfo{volume}{73}},
  \bibinfo{pages}{033312} (\bibinfo{year}{2006}).

\bibitem[{\citenamefont{Wang et~al.}(2007)\citenamefont{Wang, Jiao, Li, Li, and
  Yan}}]{Wan07125416}
\bibinfo{author}{\bibfnamefont{S.-K.} \bibnamefont{Wang}},
  \bibinfo{author}{\bibfnamefont{H.}~\bibnamefont{Jiao}},
  \bibinfo{author}{\bibfnamefont{F.}~\bibnamefont{Li}},
  \bibinfo{author}{\bibfnamefont{X.-Q.} \bibnamefont{Li}}, \bibnamefont{and}
  \bibinfo{author}{\bibfnamefont{Y.~J.} \bibnamefont{Yan}},
  \bibinfo{journal}{Phys. Rev. B} \textbf{\bibinfo{volume}{76}},
  \bibinfo{pages}{125416} (\bibinfo{year}{2007}).

\bibitem[{\citenamefont{Foros et~al.}(2005)\citenamefont{Foros, Brataas,
  Tserkovnyak, and Bauer}}]{For05016601}
\bibinfo{author}{\bibfnamefont{J.}~\bibnamefont{Foros}},
  \bibinfo{author}{\bibfnamefont{A.}~\bibnamefont{Brataas}},
  \bibinfo{author}{\bibfnamefont{Y.}~\bibnamefont{Tserkovnyak}},
  \bibnamefont{and} \bibinfo{author}{\bibfnamefont{G.~E.} \bibnamefont{Bauer}},
  \bibinfo{journal}{Phys. Rev. Lett.} \textbf{\bibinfo{volume}{95}},
  \bibinfo{pages}{016601} (\bibinfo{year}{2005}).

\bibitem[{\citenamefont{Covington et~al.}(2004)\citenamefont{Covington,
  AlHajDarwish, Ding, Gokemeijer, and Seigler}}]{Cov04184406}
\bibinfo{author}{\bibfnamefont{M.}~\bibnamefont{Covington}},
  \bibinfo{author}{\bibfnamefont{M.}~\bibnamefont{AlHajDarwish}},
  \bibinfo{author}{\bibfnamefont{Y.}~\bibnamefont{Ding}},
  \bibinfo{author}{\bibfnamefont{N.~J.} \bibnamefont{Gokemeijer}},
  \bibnamefont{and} \bibinfo{author}{\bibfnamefont{M.~A.}
  \bibnamefont{Seigler}}, \bibinfo{journal}{Phys. Rev. B}
  \textbf{\bibinfo{volume}{69}}, \bibinfo{pages}{184406}
  (\bibinfo{year}{2004}).

\end{thebibliography}

\end{document}